\documentclass[aps, prb, twocolumn, showpacs, superscriptaddress]{revtex4-1}

\usepackage{amsmath,amssymb}
\usepackage{graphicx}
\usepackage{dcolumn}
\usepackage{bm}
\begin{document}

\title {Alternative Kondo breakdown mechanism: Orbital-selective orthogonal metal transition}
\author{Yin Zhong}
\email{zhongy05@hotmail.com}
\affiliation{Center for Interdisciplinary Studies $\&$ Key Laboratory for
Magnetism and Magnetic Materials of the MoE, Lanzhou University, Lanzhou 730000, China}
\author{Ke Liu}
\affiliation{Institute of Theoretical Physics, Lanzhou University, Lanzhou 730000, China}
\author{Yong-Qiang Wang}
\affiliation{Institute of Theoretical Physics, Lanzhou University, Lanzhou 730000, China}
\author{Hong-Gang Luo}
\email{luohg@lzu.edu.cn}
\affiliation{Center for Interdisciplinary Studies $\&$ Key Laboratory for
Magnetism and Magnetic Materials of the MoE, Lanzhou University, Lanzhou 730000, China}
\affiliation{Beijing Computational Science Research Center, Beijing 100084, China}


\begin{abstract}
In a recent paper of Nandkishore, Metlitski and Senthil [arXiv:cond-mat/1201.5998v2 (2012)], a concept of orthogonal metal has been introduced to reinterpret the disordered state of slave-spin representation in the Hubbard model as an exotic gapped metallic state. We extend this concept to study the corresponding quantum phase transition in the extended Anderson lattice model. It is found that the disordered state of slave spins in this model is an orbital-selective orthogonal metal, a generalization of the concept of the orthogonal metal in the Hubbard model. The quantum critical behaviors are multiscale and dominated by a $z=3$ and $z=2$ critical modes in high and low temperature regime, respectively. Such behaviors are obviously in contrast to the naive expectation in the Hubbard model. The result provides alternative Kondo breakdown mechanism for heavy fermion compounds underlying the physics of the orbital-selective orthogonal metal in the disordered state, which is different from the conventional Kondo breakdown mechanism with the fractionalized Fermi liquid picture. This work is expected to be useful in understanding the quantum criticality happening in some heavy fermion materials and other related strongly correlated systems.
\end{abstract}

\maketitle

\section{Introduction} \label{intr}
Understanding the elusive non-Fermi liquid and its corresponding quantum criticality is one of central issues in modern condensed matter
physics.\cite{Sachdev2011,Sachdev2003,Rosch,Sachdev2008,Gegenwart,Si,Powell} To attack this challenging problem, one popular idea is to fractionalize the electrons in the model Hamiltonian into more elementary collective excitations, namely, quasiparticles like spinon, holon, and so on, near the putative quantum critical points due to wild quantum fluctuations.\cite{Wen,Senthil2003,Senthil2004,Senthil3,Senthil4,Florens,Lee2005,Pepin2005,Kim2006,Senthil2008,Kim2010,Senthil2010}

Generally, fractionalization can be performed in terms of many slave-particle theories, specifically,
for Hubbard model there exist slave bosons, slave rotors, slave spins, etc.\cite{Kotliar,Florens,Lee2005,Senthil2008,deMedici,Hassan,Ruegg,Yu,Nandkishore} In recent years, the slave spins approach has attracted much interest in the study of multi-orbital Hubbard model since it is easier to formulate than the conventional slave boson or slave rotor techniques, \cite{deMedici,Hassan,Ruegg,Yu} which is due to the fact that the slave spin
approach only has a minimum gauge structure, namely, $Z_{2}$ symmetry, \cite{deMedici,Ruegg,Nandkishore} while the other two theories have a $U(1)$ gauge symmetry. In addition, whether the $U(1)$ gauge theory in (2+1)d is confined or not is still hot debated but the deconfinement of $Z_{2}$ gauge theory in (2+1)d is undisputed.\cite{Kogut,Polyakov1975,Polyakov1977,Nagaosa,Herbut2003,Hermele,Kim,Lee2008,Zhong} Due to these advantages of the slave spin representation, this representation has been employed to study the Mott (for single band) or the orbital-selective Mott( for multi-orbital) transitions of the Hubbard model. \cite{deMedici,Hassan,Ruegg,Yu} The Mott transition has been identified as the slave spin ordering or disordering transitions. \cite{deMedici,Ruegg}

However, very recently, Nandkishore, Metlitski and Senthil\cite{Nandkishore} reinspected the slave spin representation of single-band Hubbard model and pointed out that the correct disordered state of slave spins is not a Mott insulator but an exotic metallic state which they called it as an orthogonal metal. This state is a compressible metal, which has the same thermodynamics and transport as the usual Landau Fermi liquid, but its electronic spectral function has a gap, thus leading to a simplest non-Fermi liquid. Therefore, there is an orthogonal metal-Fermi liquid transition instead of Mott transition in the slave spin representation of the single-band Hubbard model.

For the case of multi-orbital models, they argue that an orbital-selective orthogonal metal (OSOM) can be identified where some orbital are factionalized to form orthogonal metal while others are still usual Fermi liquid.\cite{Nandkishore} Therefore, it leads to an orbital-selective orthogonal metal transition but not a previously expected orbital-selective Mott transition for the slave spin representation of the multi-orbital Hubbard model.

Meanwhile, it is noted that the quantum phase transition (QPT) of many heavy fermions compounds is also modeled in terms of a two-orbital model, namely, the Anderson/Kondo lattice model, which describes one strongly correlated band hybridizing with another non-correlated band. Since this is indeed a multi-orbital model and may be highly relevant to quantum criticality observed in experiments, \cite{Rosch,Gegenwart,Si,Vojta,Custers1,Custers2} it is interesting to study whether an orbital-selective orthogonal metal and the corresponding orbital-selective transition really exist. Moreover, it is also desirable to see whether and how this possible new transition relate to the well-studied Kondo breakdown mechanism which is proposed to explain elusive non-Fermi liquid states and quantum critical behaviors in some heavy fermions materials. \cite{Senthil2003,Senthil2004,Paul,Pepin,Pepin2008,Paul2008,Kim2010,Vojta}

At the first sight, it seems to be straightforward to do it. However, different from the multi-orbital Hubbard model, a hybridization term, which is a new feature for Anderson lattice-like models responsible for the celebrated Kondo effect, exists between two distinct bands and could lead to totally different critical behaviors. We find that this hybridization term is crucial for the QPT and it leads to a multiscale quantum critical behavior like standard Kondo breakdown mechanism instead of a $\varphi^{4}$ criticality with $z=1$ expected from naive application of multi-orbital Hubbard models. Comparison between our treatment and the standard Kondo breakdown mechanism is also discussed. Moreover, we confirm the disordered state of slave spins is indeed an OSOM as expected by Nandkishore, Metlitski and Senthil \cite{Nandkishore} and the corresponding phase transition is also the expected orbital-selective orthogonal metal transition. As byproduct, we have also constructed a path integral formulism for the $Z_{2}$ slave-spin representation of the extended Anderson lattice model.

The remainder of this paper is organized as follows. In Sec.\ref{sec2}, we introduce an extended Anderson lattice
model and reformulated it in terms of the slave spin representation. Meanwhile, a useful path integral formulism is also constructed in this section. Then, a mean-field decoupling is used and two resulting mean-field states are analyzed in Sec.\ref{sec3}. One state is the expected orbital-selective orthogonal metal and the other is the usual heavy Fermi liquid. In Sec.\ref{sec4}, the QPT between these two states are discussed and an effective theory is used to clarify the correct critical behaviors, which leads to an identification of a multiscale criticality. In Sec.\ref{sec5}, we compare the results from the orbital-selective orthogonal metal transition to the usual Kondo breakdown mechanism. Finally, Sec.\ref{sec6} is devoted to a concise conclusion.

\section{$Z_{2}$ slave spin representation and extended Anderson lattice model} \label{sec2}
The model we used is an extended Anderson lattice model, \cite{Nandkishore,Paul,Pepin,Pepin2008,Paul2008}
\begin{eqnarray}
&&H=-\sum_{ij\sigma}t_{ij}d_{i\sigma}^{\dag}d_{j\sigma}+\sum_{ij}V_{ij}n_{i}n_{j}+\sum_{i}(\varepsilon_i - \mu)n_{i}\nonumber\\
&&-\sum_{ij\sigma}g_{ij}c_{i\sigma}^{\dag}c_{j\sigma}-\mu\sum_{i\sigma}c_{i\sigma}^{\dag}c_{i\sigma}
+V\sum_{i\sigma}(d_{i\sigma}^{\dag}c_{i\sigma}+h.c.), \label{eq1}
\end{eqnarray}
where $n_{i}=\sum_{\sigma}d_{i\sigma}^{\dag}d_{i\sigma}$, $\varepsilon_i$ is the on-site energy, $\mu$ is
the chemical potential, $V_{ij}$ is the Coulomb interaction which includes the on-site energy $V_{ii}=U/2$ when $i=j$, $t_{ij}$ and $g_{ij}$ are hopping integrals for localized electron $d_{i\sigma}$ and conducting electron $c_{i\sigma}$, respectively. $V$ is the hybridization between the $d$ and $c$ bands. This model describes a strongly correlated band $d$ hybridizing ($V$) with another non-correlated band $c$ and is expected to capture basic properties of some heavy
fermions systems, particularly the ones near quantum critical points. Since we expect there is at least a QPT (particularly related to non-Fermi liquid behaviors) in this model, it is natural and helpful to use techniques of fractionalization which is ideal to achieve some non-Fermi liquid behaviors and the corresponding QPT. Here, motivated by recent intensive studies of slave spin representation of the Hubbard model, we will also use the slave spin technique in our discussion of the above extended Anderson lattice model.

\subsection{$Z_{2}$ slave-spin representation}
In the treatment of $Z_{2}$ slave spin approach, the local electron $d_{\sigma}$ is fractionalized into a new auxiliary fermion $f_{\sigma}$ and a slave spin $\tau^{x}$ as\cite{deMedici,Ruegg}
\begin{equation}
d_{i\sigma}=f_{i\sigma}\tau_{i}^{x}\label{eq2}
\end{equation}
with a constraint $\tau_{i}^{z}=-(1-2f_{i\uparrow}^{\dag}f_{i\uparrow})(1-2f_{i\downarrow}^{\dag}f_{i\downarrow})$ enforced in every site. Under this representation, the original Hamiltonian can be reformulated as
\begin{eqnarray}
&&H=-\sum_{ij\sigma}t_{ij}\tau_{i}^{x}\tau_{j}^{x}f_{i\sigma}^{\dag}f_{j\sigma}+\sum_{ij}V_{ij}n_{i}^{f}n_{j}^{f}+\sum_{i\sigma}(\varepsilon_i-\mu)f_{i\sigma}^{\dag}f_{i\sigma}\nonumber\\
&&-\sum_{ij\sigma}g_{ij}c_{i\sigma}^{\dag}c_{j\sigma}-\mu\sum_{i\sigma}c_{i\sigma}^{\dag}c_{i\sigma}
+V\sum_{i\sigma}(f_{i\sigma}^{\dag}\tau_{i}^{x}c_{i\sigma}+h.c.).\label{eq3}
\end{eqnarray}
It is noted that the first three terms of this Hamiltonian are just the slave spin representation of Hubbard model and have been intensively studied by many authors.\cite{deMedici,Hassan,Ruegg,Yu,Nandkishore}
While most of these authors identified the quantum phase transition between the slave spin disorder or order as a Mott or orbital-selective Mott transition, Nandkishore, Metlitski and Senthil\cite{Nandkishore} recently pointed out that the disordered state of slave spins is not a Mott insulator but an exotic metallic state namely, the orthogonal metal. Thus, there is orthogonal metal-Fermi liquid transition instead of Mott transition in the slave spin representation of Hubbard model. Following the same methodology, we may expect that an orbital-selective orthogonal transition could be found in the extended Anderson lattice model since we are treating a two-band model where one band has strong correlation between local electrons while the other is basically a free Fermi gas.

Before leaving this subsection, we should emphasize that although the physical $d$ electron has been fractionalized into an auxiliary fermion $f_{\sigma}$ and a slave spin $\tau^{x}$, the quantum number of the electron (the spin-$\frac{1}{2}$ and the charge $e$) are both carried by the $f$ fermion, which is quite different from slave boson or slave rotor approaches where the charge and spin degree of freedom are solely carried by bosonic particles and fermionic spinons, respectively. This point is not noticed until the recent interesting work of Nandkishore, Metlitski and Senthil\cite{Nandkishore} but has crucial
influence on the correct interpretation of the disordered state of the slave spin. As argued by Nandkishore, Metlitski and Senthil, a U(1) rotation of physical electron $d$ can only be matched by a U(1) rotation of $f$ fermion while the slave spin $\tau^{x}$ do not change because it is purely real. Therefore electric charge must be only carried by $f$ fermion but not the slave spin since it corresponds to the Noether charge of the U(1) symmetry. Hence, it can be a metallic state even if the slave spin is gapped when the $f$ fermions form Fermi liquid. In the paper of Nandkishore, Metlitski and Senthil, \cite{Nandkishore} they named it as orthogonal metal to emphasize that it is metallic indeed and not a misunderstood Mott insulator. As a result, the phase transition in the $Z_{2}$ slave spin theory of Hubbard is orthogonal metal-Fermi liquid transition in replace of the Mott transition.

\subsection{Path integral formulism for the $Z_{2}$ slave-spin representation of the extended Anderson lattice model}
Before turning to discuss the mean-field treatment, we present the construction of path integral for $Z_{2}$ slave-spin approach of the extended Anderson lattice model in this subsection. To this aim, we follow the approach of Ref. [\onlinecite{Senthil2000}] where the general $Z_{2}$ gauge theory is constructed in an extended Hubbard model.

The construction of path integral is to calculate the partition function $Z=\text{Tr}(e^{-\beta \hat{H}}\hat{P})$ where
$\hat{P}$ is the projective operator to exclude unphysical states introduced by $Z_{2}$ slave-spin representation.
Here we use
\begin{equation}
\hat{P}=\prod_{i}\frac{1}{2}(1+(-1)^{\frac{1}{2}[\tau_{i}^{z}+1-2(n_{i}^{f}-1)^{2}]}).\label{eq4}
\end{equation}
This choice has the advantage to meet the mean-field theory of $Z_{2}$ slave-spin approach. Obviously, one can employ another equivalent projective operator\cite{Ruegg2012}
\begin{equation}
\hat{P}=\prod_{i}\frac{1}{2}(1+(-1)^{\frac{1}{2}[\tau_{i}^{z}-1+2n_{i}^{f}]}).\label{eq5}
\end{equation}
We will use the first definition of $\hat{P}$ in the following discussion.
Follow Ref. [\onlinecite{Senthil2000}], the projective operator can be reformulated by introducing auxiliary Ising field $\sigma_{i}=\pm1$
\begin{equation}
\hat{P}=\prod_{i}\frac{1}{2}\sum_{\sigma_{i}=\pm1}e^{i\frac{\pi}{4}(\sigma_{i}-1)[\tau_{i}^{z}+1-2(n_{i}^{f})^{2}]}.\label{eq6}
\end{equation}
Since $[\hat{P},H]=0$, one can define an effective Hamiltonian $H_{eff}$ as
\begin{eqnarray}
H_{eff}=H+\sum_{i}i\frac{\pi}{4}(1-\sigma_{i})[\tau_{i}^{z}+1-2(n_{i}^{f})^{2}].\label{eq7}
\end{eqnarray}
Then using the same method in the treatment of quantum Ising model (see Appendix A) and standard coherent state representation of fermions, one obtains the path integral formulism of $Z_{2}$ slave-spin representation of the extended Anderson lattice model
\begin{equation}
Z=\prod_{i}\int d\bar{f}_{i}df_{i}d\varphi_{i}\delta(\varphi^{2}_{i}-1)d\sigma_{i}\delta(\sigma^{2}_{i}-1) e^{-S}\label{eq8}
\end{equation}
and
\begin{eqnarray}
&&S=\int d\tau[\sum_{i\sigma}\bar{f}_{i\sigma}(\partial_{\tau}+\varepsilon_{i}-\mu)f_{i\sigma}+\sum_{ij}V_{ij}n_{i}^{f}n_{j}^{f}  \nonumber\\
&&\hspace{1cm} +\sum_{i}\frac{1}{2}(\partial_{\tau}\varphi_{i})^{2}-\sum_{ij\sigma}t_{ij}\varphi_{i}\varphi_{j}\bar{f}_{i\sigma}f_{j\sigma} \nonumber\\
&&\hspace{1cm} +\sum_{ij\sigma}\bar{c}_{i\sigma}(-g_{ij}-\mu\delta_{ij})c_{i\sigma}+V\sum_{i\sigma}\varphi_{i}(\bar{f}_{i\sigma}c_{i\sigma}+c.c.) \nonumber\\
&&\hspace{1cm} +\sum_{i}i\frac{\pi}{4}(1-\sigma_{i})[1-2(n_{i}^{f})^{2}]],\label{eq9}
\end{eqnarray}
where we have used $\tau_{i}^{x}|\varphi\rangle=\varphi_{i}|\varphi\rangle$ with $\varphi=\pm1$ and $\tau_{i}^{z}|\varphi\rangle
=|\varphi_{1}\rangle|\varphi_{2}\rangle|\varphi_{3}\rangle\cdot\cdot\cdot|-\varphi_{i}\rangle\cdot\cdot\cdot|\varphi_{N}\rangle$ to avoid confusion with auxiliary Ising field $\sigma_{i}$. The above action is our main result in this subsection and further approximations have to be made in order to gain some physical insights.

\section{mean-field theory and Orbital-selective orthogonal metal} \label{sec3}
Undoubtedly, it is a formidable task to treat the Hamiltonian of slave-spin representation of extended Anderson lattice model (Eq.\ref{eq3}) exactly, thus, here we only consider a mean field treatment and will reinclude the fluctuation effect in the discussion of critical properties of the next section.

It is straightforward to derive a mean field Hamiltonian as follows\cite{Nandkishore}
\begin{eqnarray}
&&H_{fc}=-\sum_{ij\sigma}\tilde{t}_{ij}f_{i\sigma}^{\dag}f_{j\sigma}+\sum_{ij}(V_{ij}+4\lambda_{i}\delta_{ij})n_{i}^{f}n_{j}^{f}\nonumber\\
&&+\sum_{i}(\varepsilon_i-\mu-\lambda_{i})n_{i}^{f}
-\sum_{ij\sigma}g_{ij}c_{i\sigma}^{\dag}c_{j\sigma}-\mu\sum_{i\sigma}c_{i\sigma}^{\dag}c_{i\sigma}\nonumber\\
&&+\tilde{V}\sum_{i\sigma}(f_{i\sigma}^{\dag}c_{i\sigma}+h.c.)\label{eq10} \\
&& H_{I}=-\sum_{ij}J_{ij}\tau_{i}^{x}\tau_{j}^{x}+\sum_{i}(\lambda_{i}\tau_{i}^{z}+\bar{V}\tau_{i}^{x})\label{eq11}
\end{eqnarray}
where the Lagrange multiplier $\lambda_{i}$ has been introduced to fulfill the constraint on average,
$\tilde{t}_{ij}={t}_{ij}\langle \tau_{i}^{x}\tau_{j}^{x}\rangle$, $\tilde{V}=V\langle\tau_{i}^{x}\rangle$, $J_{ij}={t}_{ij}\sum_{\sigma}\langle f_{i\sigma}^{\dag}f_{j\sigma}\rangle + c.c.$, $\bar{V}=V\sum_{\sigma}\langle f_{i\sigma}^{\dag}c_{i\sigma}\rangle + c.c.$.
The decoupled Hamiltonian $H_{I}$ is a generalized transverse Ising model and $H_{fc}$ describes $f$ fermions hybridizing with the conducting electrons.

Let us first focus on the quantum Ising model (Eq.\ref{eq11}). It is well known that the standard transverse Ising model
in one spatial dimension can be exactly solved by Jordan-Wigner transformation and it has two phases with the critical
exponents being the same as two-dimensional classical Ising model.\cite{Sachdev2011} Beyond one spatial dimension, to our knowledge, no exact solutions exist for the quantum Ising model until now. However, one may define $\langle \tau^{x}\rangle $ as a useful order parameter and there are at least two phases in two and three space dimensions. (It is just this case in the study of single-band Hubbard model in terms of some mean-field approximations and the Schwinger bosons theory.\cite{deMedici,Ruegg,Nandkishore}) One is a magnetic ordered state with $\langle \tau^{x}\rangle \neq0$ while the other is described by a vanished $\langle \tau^{x}\rangle$ and is a disordered state with an excitation gap. Moreover, there must be a quantum critical point (QCP), whose critical properties could be described by a quantum $\varphi^{4}$ theory, between these two distinct phases.

In the case of the above generalized quantum Ising model, we may simply assume that it has a magnetic ordered ($\langle \tau^{x}\rangle \neq0$) and disordered ($\langle \tau^{x}\rangle =0$) states with a QCP between them to simplify our treatment. (Readers can refer to Appendix A for such $\varphi^{4}$ theory.)

Next, we treat the Hamiltonian $H_{fc}$. It is noted that $H_{fc}$ is still an interacting Hamiltonian thus has many
possible phases. Because usually it is more interesting to study the instability of Landau Fermi liquid to other states via quantum phase transitions, we here assume $f$ fermions form a Fermi liquid and have a sharply defined Fermi surface in the remaining parts of the present paper.

Under these assumptions, it is safe to discard the interacting terms between $f$ fermions since Landau Fermi liquid is basically a non-interacting gas and we will use a modified Hamiltonian $\tilde{H}_{fc}$, which reads
\begin{eqnarray}
&&\tilde{H}_{fc}=-\sum_{ij\sigma}\tilde{t}_{ij}f_{i\sigma}^{\dag}f_{j\sigma}+\sum_{i}(\varepsilon_i-\mu-\lambda_{i})n_{i}^{f}
-\sum_{ij\sigma}g_{ij}c_{i\sigma}^{\dag}c_{j\sigma}\nonumber\\
&&-\mu\sum_{i\sigma}c_{i\sigma}^{\dag}c_{i\sigma}
+\tilde{V}\sum_{i\sigma}(f_{i\sigma}^{\dag}c_{i\sigma}+h.c.).\label{eq12}
\end{eqnarray}
The above Hamiltonian can be readily diagonalized in the momentum space. If the renormalized hybridization $\tilde{V}$ is not zero ($\langle \tau^{x}\rangle \neq0$), one obtains two new bands, namely,  $E_{\pm}(k)=\frac{1}{2}(\tilde{\varepsilon}_{k}+g_{k}\pm\sqrt{(\tilde{\varepsilon}_{k}-g_{k})^{2}+4\tilde{V}^{2}})$ where $\tilde{\varepsilon}_{k},g_{k}$ are single-particle energy for $f$ and $c$ fermions, respectively. This is just the heavy fermion band found in the more conventional slave boson theory of Anderson or Kondo lattice
models. One may interpret the case with non-zero effective hybridization $\tilde{V}$ as a heavy Fermi liquid.\cite{Senthil2003,Senthil2004,Paul,Pepin,Pepin2008,Paul2008,Vojta}
In fact, based on the discussion in the slave boson theory,\cite{Senthil2003,Senthil2004,Paul,Pepin,Pepin2008,Paul2008,Vojta} a nonzero $\tilde{V}$ also signals a development of Kondo effect. Meanwhile, to further confirm whether this state ($\tilde{V}$) is a Fermi liquid or not, it is helpful to inspect the behavior of the quasiparticle, particularly, its single-particle Green's or equivalently its spectral function.

The Green's function of the physical $d$ (localized) electrons is defined as\cite{deMedici,Ruegg,Nandkishore}
\begin{eqnarray}
G_{d\sigma}(i,j,t)&&=-i\langle T d_{j\sigma}(t)d_{i\sigma}^{\dag}(0)\rangle\nonumber\\
&&=-i\langle T \tau_{j}^{x}(t)\tau_{i}^{x}(0)f_{j\sigma}(t)f_{i\sigma}^{\dag}(0)\rangle.\label{eq13}
\end{eqnarray}
 At the mean field level, the $f$ fermion and slave spin are decoupled from each other and the resulting Green's function can be written as\cite{deMedici,Ruegg,Nandkishore}
\begin{equation}
G_{d\sigma}(i,j,t)\approx G_{spin}(i,j,t)G_{f\sigma}(i,j,t)\label{eq14}
\end{equation}
where we have define $G_{spin}(i,j,t)=\langle T\tau_{j}^{x}(t)\tau_{i}^{x}(0)\rangle $, $G_{f\sigma}(i,j,t)=-i\langle T f_{j\sigma}(t)f_{i\sigma}^{\dag}(0)\rangle $. For the free $f$ fermion, its Green's function and spectral function $A_{f\sigma}(k,\omega)$ can be easily found as  $G_{f\sigma}(k,\omega)=\frac{1}{\omega-\tilde{\varepsilon}_{k}+i\delta}$ and $A_{f\sigma}(k,\omega)=\delta(\omega-\tilde{\varepsilon}_{k})$, respectively.

In the case of $\tilde{V}\neq0$ (ordered state of the corresponding quantum Ising model),
the Green's function of $d$ electron has the form\cite{Nandkishore}
\begin{equation}
G_{d\sigma}(k,\omega)=\frac{\langle \tau^{x}\rangle^{2}}{\omega-\tilde{\varepsilon}_{k}+i\delta}\label{eq15}
\end{equation}
and its spectral function is
\begin{equation}
A_{d\sigma}(k,\omega)=\langle \tau^{x}\rangle^{2}\delta(\omega-\tilde{\varepsilon}_{k}).\label{eq16}
\end{equation}
The above spectral function with a non-zero $Z=\langle\tau^{x}\rangle^{2}$ is a key characteristic of the Landau Fermi liquid thus we conclude the whole system is a conventional Fermi liquid when the slave spin is in its ordered state.\cite{Nandkishore} In addition, due to the intensive renormalization effect (two new bands with a large effective mass being inversely proportional to $Z$ approximately), this Fermi liquid can be identified as heavy Fermi liquid as what has been done in the slave boson treatment of Anderson and Kondo lattice models.\cite{Senthil2003,Senthil2004,Paul,Pepin,Pepin2008,Paul2008,Vojta}

\subsection{Orbital-selective orthogonal metal in the extended Anderson lattice model}
In contrast, for a vanished $\tilde{V}$ (disordered state of the slave spin), the two flavors of fermions decouple from each other and at the same time, the slave spin will acquire an excitation gap. This can be seen as follows. In the low energy limit, the
quantum Ising model can be described by an effective $\varphi^{4}$ theory where space and time are on an equal footing and we obtain the Green's function of slave spin as follows\cite{Nandkishore}
\begin{equation}
G_{spin}(k,\omega)\sim\frac{1}{\Delta^{2}+k^{2}-(\omega+i\delta)^{2}}.\label{eq17}
\end{equation}
And the corresponding spectral function can also be easily derived with the form
\begin{equation}
A_{spin}(k,\omega)=\delta(\omega^{2}-(\Delta^{2}+k^{2})).\label{eq18}
\end{equation}
Clearly, the spectral function of the slave spin has an excitation gap $\Delta$. Therefore, the $d$ electron will also acquire a gap with $Z=0$. According to the definition of the orthogonal metal in the paper of Nandkishore, Metlitski and Senthil,\cite{Nandkishore} if a state has a gap for single-particle excitation and the same thermodynamics and transport properties as Landau Fermi liquid, it could be identified as an orthogonal metal. In our case, the physical $d$ electron has a excitation gap while the $f$ fermions form Fermi gas. Most importantly, the $f$ fermions carry both charge and spin degrees of freedom of the physical $d$ electrons, thus $f$ fermions will contribute to the thermodynamics, charge and spin transports exactly in
the same way as real electrons.(The contribution of slave spins can be neglected in the low energy limit, since they are gapped in the disordered state.) Therefore, the $d$ electrons are in an orthogonal metal and the whole system are consisted of a Fermi liquid of $c$ electrons and an orthogonal metal of $d$ electrons which are mutually decoupled in mean-field level. (This is still true when fluctuations are included since the slave spin is in its disordered state and has a gap.) According to Nandkishore, Metlitski and Senthil,\cite{Nandkishore} this state is an orbital-selective orthogonal metal (OSOM).

We note the properties of QCP between these two states have not been discussed in this section, since approaching the critical point, fluctuation effect may dominate and here we will leave its discussion in the next section. One will find that the critical fluctuation of the slave spin indeed dominate the thermodynamics and transport behaviors in the
QCP and in the quantum critical regime.

\section{Quantum phase transition from orbital-selective orthogonal metal to heavy Fermi liquid} \label{sec4}
Having analyzed the properties of the heavy Fermi liquid and a metallic state combining both an orthogonal metal state for $d$ electrons and a Fermi liquid for conducting electrons, in this section, we proceed to discuss the phase transition between these two distinct states.

First, it is useful to refine how the heavy Fermi liquid get heavy and die approaching the QCP.\cite{Coleman}
In the previous section, we have obtained the spectral function of physical $d$ electrons as
\begin{equation}
A_{d\sigma}(k,\omega)=Z\delta(\omega-\tilde{\varepsilon}_{k})\label{eq19}
\end{equation}
where the quasiparticle spectral weight $Z=\langle\tau^{x}\rangle^{2}$.\cite{Nandkishore} Since $\langle\tau^{x}\rangle$ is the order parameter of the quantum Ising model (Eq.\ref{eq11}), it vanishes as $(1-h/h_{c})^{\beta}$ ($h_{c}$ is the critical field of the Ising model and $\beta$ is the critical exponent) when approaching the QCP. Therefore, the quasiparticle spectral weight of $d$ electrons has to vanish as
\begin{equation}
Z\sim(1-h/h_{c})^{2\beta}\label{eq20}
\end{equation}
if one approaches the QCP from the heavy Fermi liquid. Thus, the effective mass of $d$ electrons may diverge as
$m^{\ast}\sim Z^{-1}\sim\frac{1}{(h_{c}-h)^{2\beta}}$ (This identification may be invalid as argued in Ref. [\onlinecite{Khodel}]) when the heavy Fermi liquid is near the critical point. Because near QCP, the effective mass of the whole system is dominated by the $d$ electrons, we expect that the effective mass of the heavy Fermi liquid will diverge in the similar way as the situation of $d$ electrons. Hence, the above argument confirms that the heavy Fermi liquids indeed get heavy approaching the QCP and ultimately die with a vanishing quasiparticle spectral weight.

Next, if one starts in the OSOM state (It is also the disordered state for the slave spin.) and approaches the QCP, the gap of $d$ electrons will vanish everywhere on a momentum space surface which should correspond to the Fermi surface of $f$ electrons. Specifically, based on the general quantum critical scaling theory, the gap may close (vanish) as\cite{Sachdev2011,Continentino}
\begin{equation}
\Delta\sim(h-h_{c})^{\nu z}\label{eq21}
\end{equation}
where $\nu$ are the critical exponent of the correlation length $\xi\sim|h-h_{c}|^{-\nu}$ and $z$ is the dynamical critical exponent ($z=1$ for Ising-like model). Thus, it is clear that at the QCP, the gap is zero and the Fermi surface of $f$ electrons starts evolving into the Fermi surface of physical $d$ electrons.

One may wonder that since the spin and charge degrees of freedom are both carried by the $f$ fermions and their Fermi surface evolve into the real Fermi surface of physical electrons, what is the slave spin and what role it plays. In our view, the slave spin represents the long-range coherence of the original model and can be considered as a neutral collective mode. This collective mode reflects the subtle topological order.\cite{Sachdev2003,Wen1991} As a matter of fact, even in the ordered state, no conventional symmetries (e.g. translation, rotation, spin rotation and so on) are broken by the condensation of the slave spin $\tau^{x}$, as also shown in the celebrated Kondo breakdown mechanism, \cite{Senthil2003,Senthil2004,Coleman2005,Paul,Pepin,Pepin2008,Paul2008,Hackl2008,Weber2008,Vojta2008,Civelli2010,Vojta,Kim2010b,Hackl2011,Benlagra2011,Tran2012} which describes a transition from a fractionalized Fermi liquid ($FL^{\ast}$) to a heavy Fermi liquid with the condensation of slave bosons but without symmetry breaking.

In what follows, we proceed to the discussion of the QCP. Naively, a reader, who is familiar with slave spin theory of Hubbard
model, may expect no new things appear in our case. However, this is not true.

In the naive expectation, one can use a $\varphi^{4}$ theory for slave spins and check that the coupling to fermions is irrelevant in the sense of renormalization group theory (RG) if the long-range Coulomb interaction is present.\cite{Nandkishore} However, in our extended Anderson lattice model (Eq.\ref{eq3}), there is an extra hybridizing term among the $f$, $c$ electrons and slave spins $\tau^{x}$. If this term is irrelevant in the low energy limit, the naive expectation will be justified, otherwise, one cannot rely on the results established in the study of Hubbard and have to analyze the critical properties specific to this model.

\subsection{RG argument on relevance/irrelevance of hybridizing term}
Here, we use a RG argument to show that the hybridizing term is relevant (marginal) in d=2 (d=3) for the fixed point
of the effective $\varphi^{4}$ theory. It can be performed as follows. At the QCP, it is useful to analyze an effective continuum quantum field theory (QFT) instead of the original lattice model. (One can find details in Appendix B.)
\begin{equation}
S=\int d^{d}xd\tau(L_{f}+L_{c}+L_{fc}+L_{I}+\cdot\cdot\cdot),\label{eq22}
\end{equation}
where
\begin{eqnarray}
&&L_{f}=\sum_{\sigma}\bar{f}_{\sigma}(\partial_{\tau}-iv\partial_{x}-vq-\frac{1}{2m}\partial_{y}^{2})f_{\sigma},\label{eq23} \\
&&L_{c}=\sum_{\sigma}\bar{c}_{\sigma}(\partial_{\tau}-iv_{0}\partial_{x}-\frac{1}{2m_{0}}\partial_{y}^{2})c_{\sigma},\label{eq24} \\
&&L_{fc}=V\sum_{\sigma}\varphi(\bar{f}_{\sigma}c_{\sigma}+c.c.),\label{eq25} \\
&&L_{I}=\frac{1}{2}[(\partial_{\tau}\varphi)^{2}+c^{2}(\partial_{y}\varphi)^{2}+r\varphi^{2}+u\varphi^{4}],\label{eq26}
\end{eqnarray}
where $v$ and $v_{o}$ are Fermi velocity for $f$ and $c$ electrons, respectively.
And $q=k_{F}-k_{F0}=\sqrt{(2\tilde{t}-\varepsilon+\mu+\lambda)/\tilde{t}}-\sqrt{(2g+\mu)/g}$ denotes the mismatch of Fermi surface for the $f$-fermion and conduction electrons with $m=1/2\tilde{t}$, $m_{0}=1/2g$ being the effective mass. The slave spin part $L_{I}$ is assumed to be described by an effective $\varphi^{4}$ theory and
other interaction terms neglected in above effective action do not interest for our purpose.\cite{Nandkishore}
If one insists that the critical point is controlled by the decoupled fixed point of the above action,
then one has $\text{dim}[f]=\text{dim}[c]=\text{dim}[\varphi]=\frac{d+1}{2}$, $\text{dim}[y]=-1$ and $\text{dim}[x]=\text{dim}[\tau]=-2$
with $z=1$. ($\text{dim}[\mathfrak{O}]$ is the scaling dimension of the quantity $\mathfrak{O}$, which is
equivalent to the statement that $\mathfrak{O'}(x',\tau')=b^{\text{dim}[\mathfrak{O}]}\mathfrak{O}(x,\tau)$ with $b$ being the scaling factor in the scaling transformation of RG.) Obviously, a scaling argument indicates the $u$ term in $L_{I}$ is irrelevant for $d>1$, thus we can neglect such term in our following treatment.

Then, using the above scaling dimensions in the tree level RG of the hybridizing term (Eq.\ref{eq26}), one obtains $\text{dim}[V]=\frac{3-d}{2}$ which means that only if $d>3$ is satisfied, the hybridizing term is indeed irrelevant. For the case of d=2, the hybridizing term is relevant and could completely destroy the free fixed point of $\varphi^{4}$ theory while the hybridizing term is marginal in case of three spatial dimensions. Although we only consider the tree level RG, the results we find have shown a sign that it is crucial to include the hybridizing term in the critical theory and the naively expected fixed point of $\varphi^{4}$ theory could not be stable when encountering the hybridizing.

Before leaving this subsection, we should emphasize that the marginal (or relevant) feature of the hybridizing term is indeed unchanged by other perturbations. For example, when we consider a system without continuous rotation symmetry (generic lattice model), there will be a Landau damping term $\delta S\propto\int_{q,\omega}(a+b|\omega|/q)|O(q,\omega)|$ with $O(x)=(\varphi(x))^{2}$ as discussed in Ref. [\onlinecite{Nandkishore}]. Obviously, this term is subleading to the hybridizing term since it involves the composite object $O(x)=(\varphi(x))^{2}$. In Ref. [\onlinecite{Nandkishore}], the authors conclude that the transition from Fermi liquid to orthogonal metal would not be described by the decoupled fixed point for generic lattice models except for a fine tuned $Z_{4}$ model. In our case, if we consider such a $Z_{4}$ model to replace of the $\varphi^{4}$ theory for slave spins, one will find again that the induced Landau damping term for composite object $O=\varphi^{2}$ is subleading to the hybridizing term. Therefore, we suspect the most relevant perturbation to the decoupled fixed point might be the hybridizing term for generic lattice models.

\subsection{Effect of the hybridizing term}
Hence, it is important to find what role the hybridizing term plays. To proceed, one can utilize the effective action Eqs. (\ref{B6}) and (\ref{B6}) derived in Appendix B instead of the one-patch action used in the RG argument.
\begin{eqnarray}
&&S=\int d\tau(L_{I}+L_{f}+L_{c}+L_{fc})\nonumber\\
&&L_{I}=\int d^{d}x\frac{1}{2}[c^{2}(\nabla\varphi)^{2}+r\varphi^{2}]\nonumber\\
&&L_{f}=\sum_{k\sigma}\bar{f}_{k\sigma}(\partial_{\tau}+\tilde{\varepsilon}(k))f_{k\sigma}\nonumber\\
&&L_{c}=\sum_{k\sigma}\bar{c}_{k\sigma}(\partial_{\tau}+g(k))c_{k\sigma}\nonumber\\
&&L_{fc}=V\int d^{d}x\sum_{\sigma}\varphi(\bar{f}_{\sigma}c_{\sigma}+c.c.).\label{eq27}
\end{eqnarray}
where $\tilde{\varepsilon}(k)=\tilde{t}k^{2}-2\tilde{t}+\varepsilon-\mu-\lambda$, $g(k)=gk^{2}-2g-\mu$ and other terms as $u\varphi^{4}$ and $(\partial_{\tau}\varphi)^{2}$ are neglected due to their irrelevance as what has been seen in previous RG argument.

A careful reader may note that the above effective action is rather similar to the one in the Kondo breakdown mechanism of Anderson/Kondo lattice model where the slave boson representation is used.\cite{Senthil2003,Senthil2004,Coleman2005,Paul,Pepin,Pepin2008,Paul2008,Hackl2008,Weber2008,Vojta2008,Civelli2010,Vojta,Kim2010b,Hackl2011,Benlagra2011,Tran2012} However, we should note that there exist some differences between these two formulism. First, in the usual Kondo breakdown mechanism, a fluctuating $U(1)$ gauge-field, which results from fluctuation over the mean-field state of spin liquid, plays an important role while no such gapless gauge-field appears in our $Z_{2}$ formulism. This is because no Heisenberg exchange term is introduced in our extended Anderson lattice model Eq. (\ref{eq1}). Second, the hybridizing effect is represented by a complex bosonic field in the Kondo breakdown mechanism and this complex bosonic field contributes to electric conduction. In contrast, in the $Z_{2}$ slave-spin representation, the slave-spin $\varphi$ denoting the hybridizing effect is a real bosonic field and it has no contributions to electric conduction at all since the electric charge is only carried by $f$-fermion though it could affect the thermal-transport.

With above distinctions in mind, following the detailed treatment of the Kondo breakdown mechanism,\cite{Pepin2008,Paul2008} we can calculate the corrections from the the hybridizing term. (We here only consider the case of three spatial dimensions for simplicity.) At the one-loop order, we have the self-energy correction for the slave-spin as
\begin{eqnarray}
&& \Pi_{fc}(k,i\Omega_{n})=\frac{2\rho_{0}}{4\pi}\int d\epsilon d\omega d\cos\theta\times\nonumber\\
&&\hspace{1cm} \left[1/((i\omega+i\Omega_{n}-\epsilon-v_{0}k\cos\theta)(i\omega-\alpha\epsilon-\alpha v_{0}q))\right.\nonumber\\
&&\hspace{2cm} \left.+(k\rightarrow-k)\right]\label{eq28}
\end{eqnarray}
where $\rho_{0}=1/2D$ ($2D$ denotes the band width of conduction electrons) is the density of state of free conduction electrons, $\alpha=v/v_{0}$ with $q$ the mismatch of two Fermi surface of $f$-fermion and conduction electron $c$. We have also used the linearized spectrums, which means $\tilde{\varepsilon}(k)\approx v(k-k_{F})$ and $g(k)\approx v_{0}(k-k_{F0})$ with $q=k_{F}-k_{F0}$. The calculation of the above integral can be found in Eq. (17) of Ref. [\onlinecite{Pepin2008}] and the only difference for our case is the existence of the second term with $k\rightarrow-k$. The appearance of such term is indeed due to the fact that the real bosonic slave-spin $\varphi$ couples $\bar{f}_{\sigma}c_{\sigma}$ and $\bar{c}_{\sigma}f_{\sigma}$ as can be seen in the last term in Eq. (\ref{eq27}).

Then, since the calculation of Eq. (\ref{eq27}) is just as Eq. (17) of Ref. [\onlinecite{Pepin2008}], we may only quote the results of Ref. [\onlinecite{Pepin2008}]. The main result is that the corrections from the hybridizing term is multiscale because of the mismatch of two Fermi surface. There exists a characteristic energy scale $E^{\star}=0.1\frac{vk_{F0}}{2}(q/k_{F0})^{3}$, above which the hybridizing fluctuation of the slave-spin is dominated by the usual Landau damping term (See Appendix C for details.)
\begin{equation}
\int d^{d}qd\omega\left(\frac{|\omega|}{q}+q^{2}+\cdots\right)|\varphi(q,\omega)|^{2}.\label{eq29}
\end{equation}
while below the characteristic energy scale $E^{\star}$, the slave-spin is not damped by the particle-hole excitation and its dynamical critical exponent is $z=2$, \cite{Pepin2008,Paul2008}
\begin{equation}
\int d^{d}qd\omega\left(-\omega+q^{2}+\cdots\right)|\varphi(q,\omega)|^{2}.\label{eq30}
\end{equation}

Obviously, the critical modes of $z=2$ and $z=3$ are more relevant than the $\varphi^{4}$ theory with $z=1$ and in the low energy limit, both thermal and transport properties will be dominated by these two kinds of critical modes.
Due to the appearance of these two modes, we could conclude that the transition from the orbital-selective orthogonal metal to heavy Fermi liquid is also multiscale as the usual Kondo breakdown mechanism. Moreover, we should emphasize that, for systems in three spatial dimensions, the above one-loop calculation (Gaussian fluctuation correction) seems sufficient both from the scaling argument and the Eliashberg analysis.\cite{Sachdev2011,Pepin2008}

\section{Comparison with usual Kondo breakdown mechanism}\label{sec5}
After the discussion in last section, we now realize that the $Z_{2}$ slave-spin representation of the extended Anderson model just provides a modified Kondo breakdown theory in compared to the standard Kondo breakdown in terms of slave bosons. However, there are some important distinctions. Our theory describes a transition
from a Fermi liquid-like OSOM state to conventional heavy Fermi liquid while there is a transition between $FL^{\ast}$ and the heavy Fermi liquid for the standard Kondo breakdown theory. Moreover, the critical fluctuation is
mainly from slave spins. In contrast, both slave bosons and gapless U(1) gauge fields contribute to the critical properties in standard one. In addition, the standard Kondo breakdown theory is usually called orbital-selective Mott transition\cite{Pepin,Pepin2008} since one of the two bands is Mott localized while in our case we find one band develops a gap but still behave as a metal (OSOM) with another band remaining a simple metal. As suggested in the paper of Nandkishore, Metlitski and Senthil, we
may also call our theory as an orbital-selective orthogonal metal transition.

Moreover, for physical observables at criticality, follow the treatment of Sec. 5 in Ref. [\onlinecite{Paul2008}], we find the specific heat behaves as $C_{v}\propto T+T^{3/2}$ (Fermi liquid-like behavior dominates.) when $T\leq E^{\star}$, while $C_{v}\propto T\log(T/E^{\star})$ dominates for $T>E^{\star}$. However, in the standard Kondo breakdown mechanism, due to the existence of the gapless $U(1)$ gauge-field, the specific heat is also dominated by the $T\log(T/E^{\star})$ term even for $T\leq E^{\star}$.

For the static spin susceptibility, the critical modes of slave-spin contributes $\delta\chi_{s}\propto-T^{2}$ ($T\leq E^{\star}$) and $\delta\chi_{s}\propto-T^{4/3}$ ($T>E^{\star}$) while the gapless $U(1)$ gauge-field gives rise to an extra contribution $\delta\chi_{s}\propto T^{2}\log(T)$ in the standard Kondo breakdown mechanism. \cite{Paul2008}

As for the crossover lines defining the quantum critical regime, our $Z_{2}$ slave-spin representation gives the same
results as standard Kondo breakdown mechanism, namely, $T\propto(r)^{2/3}$ ($T\leq E^{\star}$) and $T\propto(r)^{3/4}$ ($T>E^{\star}$) with $r^{2}$ is proportional to the gap of slave spins in the disordered states.\cite{Pepin2008,Paul2008} Meanwhile, the temperature dependent resistivity of $Z_{2}$ slave-spin representation recovers the same behavior ($\delta\rho(T)\propto T^{2}$ for $T\leq E^{\star}$ and $\delta\rho(T)\propto T\log(T/E^{\star})$ when $T>E^{\star}$.) in the usual Kondo breakdown mechanism.\cite{Pepin2008,Paul2008}

The high temperature results ($T>E^{\star}$) of Wiedemann-Franz ratio and Gr\"{u}neisen ratio in our present theory
are identical to the one in well-studied Kondo breakdown mechanism.\cite{Kim-K2008,Kim-K2009} However, due to the lack of the gapless $U(1)$ gauge-field, in the low temperature regime, we find a constant Gr\"{u}neisen ratio and the temperature dependent part of the Wiedemann-Franz ratio behaves as $\delta L(T)\propto\frac{v}{v_{0}}$ instead of $\delta L(T)\propto\frac{T^{5/3}v}{T^{2}v_{0}}$ in the usual Kondo breakdown mechanism.\cite{Kim-K2008,Kim-K2009}

We should remind the reader that although the critical behaviors are controlled by the Gaussian theory with $z=3$ or $z=2$ but not by the $\varphi^{4}$ theory which has $z=1$,
the results of previous sections are all unchanged since the slave spins are gapped or condensed which cannot modify
the qualitative properties of OSOM and heavy Fermi liquid. However, the critical exponents have to be mean-field-like because in our theory $d+z>d_{c}=4$ where $d_{c}$ is the upper critical dimension. But, we also note that for d=2, there is still a debate on whether the critical behaviors are really Gaussian.\cite{Lee2009,Metlitski,Mross2010} This debate results from
that the critical theory is found to be strongly correlated in spite of the seeming Gaussian feature and no controllable techniques in the framework of conventional QFT can be used to extract useful observables.\cite{Lee2009,Metlitski,Mross2010} Therefore, some researches are turning into the studies of the derivative of strings theory,\cite{Herzog,Lee2009d,Sachdev3,Sachdev2012} the anti-de Sitter/conformal field theory (AdS/CFT) correspondence (or gauge/gravity duality)\cite{Hartnoll,McGreevy} and hope this new mathematical machinery may help us to inspect difficult condensed matter problems from the viewpoint of quantum gravity. Besides, we expect that the exotic orthogonal metal or its derivative, the orbital-selective orthogonal metal, may be realized in the
powerful gauge/gravity duality via confining geometries that terminate along the holographic $\mu$ direction as the realization of usual Fermi liquid \cite{Sachdev2012} but with carefully chosen boundary conditions which are needed to be uncovered in near future.

\section{Conclusion}\label{sec6}
In the present paper we have shown that an orbital-selective orthogonal metal state can exist in the slave spin representation of an extended Anderson lattice model. The corresponding transition is identified as an orbital-selective orthogonal metal transition as expected (Recently, we note that a U(1) slave-spin representation has been proposed and it indeed describes a Mott transition in contrast to the $Z_{2}$ case employed in this paper.\cite{Yu2012}). However, in contrast to the naive expectation based on the insight from Hubbard model, a multi-scale quantum critical behavior is found in stead of a $z=1$ $\varphi^{4}$ critical theory of the Hubbard model, due to the highly relevant hybridization term appearing in the Anderson lattice-like models.Moreover, it is noted that the result we obtained is rather similar to the standard Kondo breakdown mechanism, \cite{Senthil2003,Senthil2004,Coleman2005,Paul,Pepin,Pepin2008,Paul2008,Hackl2008,Weber2008,Vojta2008,Civelli2010,Vojta,Kim2010b,Hackl2011,Benlagra2011,Tran2012} thus we may consider it as alterative Kondo breakdown mechanism with an orbital-selective orthogonal metal transition in stead of an
orbital-selective Mott transition of the standard one. However, we also note there exist some interesting differences between these two formulism since no gapless $U(1)$ gauge fluctuation appears near criticality in our $Z_{2}$ slave-spin representation. We hope that our present work could be useful for further study in the quantum criticality happening in some heavy fermions materials and other related strongly correlated systems.

\begin{acknowledgments}
We thank professor Ki-Seok Kim for careful reading and useful discussions. The work was supported partly by NSFC, the Program for NCET, the Fundamental Research Funds for the Central Universities and the national program for basic research of China.
\end{acknowledgments}

\appendix
\section{path integral for the quantum Ising model in transverse field}
The quantum Ising model in transverse field is defied as\cite{Sachdev2011}
\begin{equation}
\hat{H}_{I}=-J\sum_{\langle ij\rangle\sigma}(\tau_{i}^{z}\tau_{j}^{z}+h.c.)-K\sum_{i}\tau_{i}^{x}\label{A1}
\end{equation}
where a ferromagnetic coupling $J>0$ is assumed and $K$ represents the the transverse external field.

At first glance, one may directly use the coherent state of spin operators in constructing the path integral representation, (One can find a brief but useful introduction to this issue in Ref. [\onlinecite{Sachdev2011}]) however, this will lead to an extra topological Berry phase term and is not easy to utilize practically. An alterative approach is to use the eigenstates of spin operator $\tau^{x}$ or $\tau^{z}$ as the basis for calculation.\cite{Stratt}
One will see this approach is free of the topological Berry phase term and give rise to a rather simple formulism.  Therefore, to construct a useful path integral representation, we will follow Ref. [\onlinecite{Stratt}].

First of all, we consider the orthor-normal basis of $N_{s}$-Ising spins as
\begin{equation}
|\sigma\rangle\equiv|\sigma_{1}\rangle|\sigma_{2}\rangle|\sigma_{2}\rangle\cdot\cdot\cdot|\sigma_{N}\rangle\label{A2}
\end{equation}
with $\sigma_{i}=\pm1$ and define
\begin{equation}
\tau_{i}^{z}|\sigma\rangle=\sigma_{i}|\sigma\rangle,\label{A3}
\end{equation}
\begin{equation}
\tau_{i}^{x}|\sigma\rangle
=|\sigma_{1}\rangle|\sigma_{2}\rangle|\sigma_{3}\rangle\cdot\cdot\cdot|-\sigma_{i}\rangle\cdot\cdot\cdot|\sigma_{N}\rangle.\label{A4}
\end{equation}
Then the partition function $Z=Tr(e^{-\beta \hat{H}})$ can be represented as
\begin{eqnarray}
Z=\sum_{\{\sigma\}=\pm1}\prod_{n=1}^{N}e^{\epsilon J\sum_{\langle ij\rangle}\sigma_{i}(n)\sigma_{j}(n)}\langle\sigma(n+1)|e^{\epsilon K\sum_{i}\tau_{i}^{x}}|\sigma(n)\rangle \nonumber
\end{eqnarray}
where $\epsilon$N=$\beta$.
The calculation of $\langle\sigma(n+1)|e^{\epsilon K\sum_{i}\tau_{i}^{x}}|\sigma(n)\rangle$ is straightforward
by exponentiating the $\tau_{i}^{x}$ matrix and one gets
\begin{eqnarray}
\langle\sigma(n+1)|e^{\epsilon K\sum_{i}\tau_{i}^{x}}|\sigma(n)\rangle&&=\frac{1}{2}(e^{\epsilon K}+e^{-\epsilon K}\sigma_{i}(n)\sigma_{i}(n+1)),\nonumber \\
&&=e^{a\sigma_{i}(n)\sigma_{i}(n+1)+b}\label{A5}
\end{eqnarray}
where $a=\frac{1}{2}[\ln\cosh(\epsilon K)-\ln\sinh(\epsilon K)]$ and $b=\frac{1}{2}[\ln\cosh(\epsilon K)+\ln\sinh(\epsilon K)]$.
Therefore, the resulting path integral formulism for the quantum Ising model in transverse field is
\begin{equation}
Z=\sum_{\{\sigma\}=\pm1}\prod_{n=1}^{N}e^{\epsilon J\sum_{\langle ij\rangle}\sigma_{i}(n)\sigma_{j}(n)+\sum_{i}a\sigma_{i}(n)\sigma_{i}(n+1)+N_{s}b}.\label{A6}
\end{equation}

Further, if one assumes the model is defined in a hyper-cubic lattice in space dimension of d, an effective theory can be derived as
\begin{equation}
Z=\int D\phi \delta(\phi^{2}-1) e^{-\int d\tau d^{d}x \frac{1}{2g}[(\partial_{\tau}\phi)^{2}+c^{2}(\nabla\phi)^{2}]},\label{A7}
\end{equation}
where $\frac{1}{2g}=(\frac{a\epsilon}{a_{0}^{d}})^{\frac{d+1}{2}}$ with $a_{0}$ being the lattice constant and $c^{2}=\frac{Ja_{0}^{d-2}}{a\epsilon}$. Moreover, in the effective theory, $\phi$ corresponds to $\tau^{z}$ while $\tau^{x}$ gives the kinetic energy term in imaginary time. Then, the standard $\phi^{4}$ theory is obtained by relaxing the hard constraint $\phi^{2}=1$ while introducing a potential energy term,
\begin{equation}
Z=\int D\phi e^{-\int d\tau d^{d}x [(\partial_{\tau}\phi)^{2}+c^{2}(\nabla\phi)^{2}+r\phi^{2}+u\phi^{4}]},\label{A8}
\end{equation}
where $r,u$ are effective parameters depending on microscopic details.

\section{Effective theory for orbital-selective orthogonal metal transition}
In this section, we would like to derive an effective theory for the orbital-selective orthogonal metal transition.
First, using the mean-field Hamiltonian Eqs. (\ref{eq11}) and (\ref{eq12}), one finds
\begin{equation}
H_{I}=-J\sum_{\langle ij\rangle}\tau_{i}^{x}\tau_{j}^{x}+\lambda\sum_{i}\tau_{i}^{z}\label{B1}
\end{equation}
\begin{eqnarray}
&&\tilde{H}_{fc}=-\tilde{t}\sum_{\langle ij\rangle\sigma}f_{i\sigma}^{\dag}f_{j\sigma}+\sum_{i}(\varepsilon-\mu-\lambda)n_{i}^{f}
-g\sum_{\langle ij\rangle\sigma}c_{i\sigma}^{\dag}c_{j\sigma}\nonumber\\
&&\hspace{2cm} -\mu\sum_{i\sigma}c_{i\sigma}^{\dag}c_{i\sigma}\label{B2}
\end{eqnarray}
where we have considering uniform $\varepsilon$, $\lambda$ and approaching the QCP from the orbital-selective orthogonal metal with only nearest-neighbor hopping for simplicity. Therefore, mean-field parameters $\bar{V}$ and $\tilde{V}$ are both zero in the above formulism. In contrast, generically, effective hopping $\tilde{t}$ and $J$ do not vanish even in the disordered state of slave spins since $\tilde{t}=\langle\tau^{x}_{i}\tau^{x}_{j}\rangle \neq\langle \tau^{x}\rangle^{2}$ as can be seen in cluster-mean-field treatment and Schwinger boson representation in Ref. [\onlinecite{Ruegg}].

Then, follow the path integral treatment for the $Z_{2}$ slave-spin representation in Sec.\ref{sec2}, we can obtain
\begin{eqnarray}
&&S_{0}=\int d\tau(L_{I}+L_{f}+L_{c})\nonumber\\
&&L_{I}=\frac{1}{2}\sum_{i}(\partial_{\tau}\varphi)^{2}-J\sum_{\langle ij\rangle}\varphi_{i}\varphi_{j}\nonumber\\
&&L_{f}=\sum_{\langle ij\rangle\sigma}\bar{f}_{i\sigma}(\partial_{\tau}-\tilde{t}+(\varepsilon-\mu-\lambda)\delta_{ij})f_{j\sigma}\nonumber\\
&&L_{c}=\sum_{\langle ij\rangle\sigma}\bar{c}_{i\sigma}(\partial_{\tau}-g-\mu\delta_{ij})c_{j\sigma}.\label{B3}
\end{eqnarray}

When one considers a square or a cubic lattice, one can find a continuum theory as
\begin{eqnarray}
&&S_{0}=\int d\tau(L_{I}+L_{f}+L_{c})\nonumber\\
&&L_{I}=\int d^{d}x\frac{1}{2g}[(\partial_{\tau}\varphi)^{2}+c^{2}(\nabla\varphi)^{2}]\nonumber\\
&&L_{f}=\sum_{k\sigma}\bar{f}_{k\sigma}(\partial_{\tau}+\tilde{\varepsilon}(k))f_{k\sigma}\nonumber\\
&&L_{c}=\sum_{k\sigma}\bar{c}_{k\sigma}(\partial_{\tau}+g(k))c_{k\sigma}.\label{B4}
\end{eqnarray}
where $\tilde{\varepsilon}(k)=\tilde{t}k^{2}-2\tilde{t}+\varepsilon-\mu-\lambda$, $g(k)=gk^{2}-2g-\mu$.

To facilitate the RG argument, it is helpful to derive a one-patch theory of the above action as what has been done in Refs. [\onlinecite{Sachdev2011}] and  [\onlinecite{Lee2009}].
\begin{eqnarray}
&&S_{0}=\int d\tau \int dx d^{d-1}y(L_{I}+L_{f}+L_{c})\nonumber\\
&&L_{I}=\frac{1}{2g}[(\partial_{\tau}\varphi)^{2}+c^{2}(\partial_{y}\varphi)^{2}]\nonumber\\
&&L_{f}=\sum_{\sigma}\bar{f}_{\sigma}(\partial_{\tau}-iv\partial_{x}-vq-\frac{1}{2m}\partial_{y}^{2})f_{\sigma}\nonumber\\
&&L_{c}=\sum_{\sigma}\bar{c}_{\sigma}(\partial_{\tau}-iv_{0}\partial_{x}-\frac{1}{2m_{0}}\partial_{y}^{2})c_{\sigma}.\label{B5}
\end{eqnarray}
Here, we have assumed a circular or sphere Fermi surface both for the $f$-fermion and conduction electrons $c$, which corresponds to the simplified quasiparticle spectrum $\tilde{\varepsilon}(k)$ and $g(k)$. $v=\sqrt{\tilde{t}(2\tilde{t}-\varepsilon+\mu+\lambda)}$ and $v_{0}=\sqrt{g(2g+\mu)}$ denotes Fermi velocity for $f$-fermion and conduction electrons, respectively. And $m=1/2\tilde{t}$, $m_{0}=1/2g$ is the effective mass. Besides, $q=k_{F}-k_{F0}=\sqrt{(2\tilde{t}-\varepsilon+\mu+\lambda)/\tilde{t}}-\sqrt{(2g+\mu)/g}$ is the mismatch of Fermi surface for the $f$-fermion and conduction electrons since the two Fermi surface need not be the same except for the case of fine tuning.

Then, one may add interaction terms into the above effective action. It is easy to see that the leading term will be the hybridizing term (See the path integral representation Eq. (\ref{eq9}).)
\begin{eqnarray}
L_{fc}=V\sum_{\sigma}\varphi(\bar{f}_{\sigma}c_{\sigma}+c.c.)\label{B6}
\end{eqnarray}
while terms as $(\partial_{y}\varphi)^{2}\bar{f}_{\sigma}f_{\sigma}$ and $\bar{f}_{\sigma}f_{\sigma}\bar{f}_{\sigma'}f_{\sigma'}$ are subleading, which come from $-\sum_{ij\sigma}t_{ij}\varphi_{i}\varphi_{j}\bar{f}_{i\sigma}f_{j\sigma}$ and the constraint term $\sum_{i}i\frac{\pi}{4}(1-\sigma_{i})[1-2(n_{i}^{f})^{2}]$, respectively.

\section{Calculation of the integral in Eq. (\ref{eq28})}
This section is devoted to the discussion of Eq. (\ref{eq28}) and we will follow the treatment of Eq. (17) in Ref. [\onlinecite{Pepin2008}].
The integral in Eq. (\ref{eq28}) reads
\begin{eqnarray}
\Pi_{fc}(k,i\Omega_{n})&&=\frac{2\rho_{0}}{4\pi}\int d\epsilon d\omega d\cos\theta\times\nonumber\\
&&[\frac{1}{(i\omega+i\Omega_{n}-\epsilon-v_{0}k\cos\theta)(i\omega-\alpha\epsilon-\alpha v_{0}q)}\nonumber\\
&&+(k\rightarrow-k)]\nonumber\\
&&=\frac{-\rho_{0}}{2v_{0}\alpha'k(1-\alpha')}([-\alpha'i\Omega_{n}+\alpha'v_{0}(k-q)]\nonumber\\
&&\times\log[-\alpha'i\Omega_{n}+\alpha'v_{0}(k-q)]\nonumber\\
&&-[-\alpha'i\Omega_{n}+\alpha'v_{0}(-k-q)]\nonumber\\
&&\times\log[-\alpha'i\Omega_{n}+\alpha'v_{0}(-k-q)]\nonumber\\
&&-[-i\Omega_{n}+\alpha'v_{0}(+k-q)]\nonumber\\
&&\times\log[-i\Omega_{n}+\alpha'v_{0}(+k-q)]\nonumber\\
&&+[-i\Omega_{n}+\alpha'v_{0}(-k-q)]\nonumber\\
&&\times\log[-i\Omega_{n}+\alpha'v_{0}(-k-q)])+(k\rightarrow-k).\nonumber
\end{eqnarray}
Here, we have defined $\alpha'=vk_{F0}\rho_{0}$. It is easy to see that the second term with $k\rightarrow-k$ contributes identically like the first lengthy term, thus the above result can be simplified as what has been done in Ref. [\onlinecite{Pepin2008}]. Due to the Fermi surface mismatch $q$, we could define a characteristic energy scale $E^{\star}=0.1\frac{vk_{F0}}{2}(q/k_{F0})^{3}$ and expand the above equation in four different regimes.

(1) $\Pi_{fc}(k,i\Omega_{n})\approx 2\rho_{0}V^{2}\log\alpha'/(1-\alpha')-4\frac{i\Omega_{n}}{\alpha'v_{0}q}$ for $k\leq q, |\Omega_{n}|\leq E^{\star}$. The dynamical critical exponent is $z=2$ and the slave-spin is an undamped bosonic mode, which corresponds to a dilute boson gas universal class.

(2) $\Pi_{fc}(k,i\Omega_{n})\approx 2\rho_{0}V^{2}\log\alpha'/(1-\alpha')+4\frac{\log|\Omega_{n}|}{\alpha'v_{0}q}$ for $k\leq q, |\Omega_{n}|\geq E^{\star}$. In this regime, the dynamical critical exponent is $z=\infty$.

(3) $\Pi_{fc}(k,i\Omega_{n})\approx 2\rho_{0}V^{2}\log\alpha'/(1-\alpha')+4\frac{\log|\Omega_{n}|}{\alpha'v_{0}k}$ for $k>q,\alpha'v_{0}k\leq|\Omega_{n}|\leq v_{0}k$, where dynamics is controlled by the critical exponent $z=1$.

(4) Finally, $\Pi_{fc}(k,i\Omega_{n})\approx 2\rho_{0}V^{2}\log\alpha'/(1-\alpha')+4\frac{|\Omega_{n}|}{\alpha'v_{0}k}$ when $k>q,|\Omega_{n}|\geq v_{0}k$. It has been shown that the spectral weight in the
($\omega,q$) space is most entirely centered in this regime with $z=3$. \cite{Pepin2008}

As discussed in Ref. [\onlinecite{Pepin2008}], at low momentum and low energy, the particle-hole continuum is gapped due to the mismatch of the two Fermi surface. Thus, there exists a propagating single boson mode with
exponent $z=2$. In contrast, at high momentum and high energy, one can safely neglect the mismatch of the two Fermi surface since the two fermion species behave as if they were identical. Therefore, the resulting polarization $\Pi_{fc}(k,i\Omega_{n})$ behaves like a usual Lindhard function with $z=3$.

It is also noted that the characteristic energy scale $E^{\star}=0.1\frac{vk_{F0}}{2}(q/k_{F0})^{3}$ would be rather small due to its dependence of $(q/k_{F0})^{3}$ with mismatch $q$ smaller than the Fermi wavevector $k_{F0}$, in general. \cite{Pepin2008} Thus, the critical behaviors will be dominated by the $z=3$ Landau damped modes at most regions of quantum critical regime except for the very low temperature region where the undamped modes dominate.


\begin{thebibliography}{99}

\bibitem{Sachdev2011}
S. Sachdev, \textit{Quantum Phase Transition}, 2nd ed. (Cambridge University Press, Cambridge, England, 2011).

\bibitem{Sachdev2003}
S. Sachdev, Rev. Mod. Phys \textbf{75}, 913 (2003).

\bibitem{Rosch}
H. V. L$\ddot{o}$hneysen, A. Rosch, M. Vojta and P. W$\ddot{o}$lfle, Rev. Mod. Phys \textbf{79}, 1015 (2007).

\bibitem{Sachdev2008}
S. Sachdev, Nature Phys \textbf{4}, 173 (2008).

\bibitem{Gegenwart}
P. Gegenwart, Q. Si and F. Steglich, Nature Phys \textbf{4}, 186 (2008).

\bibitem{Si}
Q. Si and F. Steglich, Science \textbf{329}, 1161 (2010).

\bibitem{Powell}
B. J. Powell and R. H. McKenzie, Rep. Prog. Phys. \textbf{74} 056501 (2011).

\bibitem{Wen}
P. A. Lee, N. Nagaosa and X. G. Wen, Rev. Mod. Phys \textbf{78}, 17 (2006).

\bibitem{Senthil2003}
T. Senthil, S. Sachdev and M. Vojta, Phys. Rev. Lett. \textbf{90}, 216403 (2003).

\bibitem{Senthil2004}
T. Senthil, M. Vojta and S. Sachdev, Phys. Rev. B \textbf{69}, 035111 (2004).

\bibitem{Senthil3}
T. Senthil et al., Science \textbf{303}, 1490 (2004).

\bibitem{Senthil4}
T. Senthil, L. Balents, S. Sachdev, A. Vishwanath and M. P. A. Fisher, Phys. Rev. B \textbf{70}, 144407 (2004).

\bibitem{Florens}
S. Florens and A. Georges, Phys. Rev. B \textbf{70}, 035114 (2004).

\bibitem{Lee2005}
S. S. Lee and P. A. Lee, Phys. Rev. Lett. \textbf{95}, 036403 (2005).

\bibitem{Pepin2005}
C. P\'epin, Phys. Rev. Lett. \textbf{94}, 066402 (2005).

\bibitem{Kim2006}
K. S. Kim, Phys. Rev. Lett. \textbf{97}, 136402 (2006).

\bibitem{Senthil2008}
T. Senthil, Phys. Rev. B \textbf{78}, 045109 (2008).

\bibitem{Kim2010}
K. S. Kim and C. L. Jia, Phys. Rev. Lett. \textbf{104}, 156403 (2010).

\bibitem{Senthil2010}
T. Grover and T. Senthil, Phys. Rev. B \textbf{81}, 205102 (2010).

\bibitem{Kotliar}
G. Kotliar and A. E. Ruckenstein, Phys. Rev. Lett. \textbf{57}, 1362 (1986).

\bibitem{deMedici}
L. de'Medici, A. Georges and S. Biermann, Phys. Rev. B \textbf{72}, 205124 (2005).

\bibitem{Hassan}
S. R. Hassan and L. de Medici, Phys. Rev. B \textbf{81}, 035106 (2010)

\bibitem{Ruegg}
A. Ruegg, S. D. Huber and M. Sigrist, Phys. Rev. B \textbf{81}, 155118 (2010).

\bibitem{Yu}
R. Yu and Q. Si, Phys. Rev. B \textbf{84}, 235115 (2011).

\bibitem{Nandkishore}
R. Nandkishore, M. A. Metlitski and T. Senthil, arXiv:cond-mat/ 1201.5998v2 (2012).

\bibitem{Kogut}
J. B. Kogut, Rev. Mod. Phys \textbf{51}, 659 (1979).

\bibitem{Polyakov1975}
A. M. Polyakov, Phys. Lett. B \textbf{59}, 82 (1975).

\bibitem{Polyakov1977}
A. M. Polyakov, Nucl. Phys. B \textbf{120}, 429 (1977).

\bibitem{Nagaosa}
N. Nagaosa, Phys. Rev. Lett. \textbf{71}, 4210 (1993).

\bibitem{Herbut2003}
I. F. Herbut, B. H. Seradjeh, S. Sachdev and G. Murthy, Phys. Rev. B \textbf{68}, 195110 (2003).

\bibitem{Kim}
K. S. Kim, Phys. Rev. B \textbf{72}, 245106 (2005).

\bibitem{Hermele}
M. Hermele, T. Senthil, M. P. A. Fisher, P. A. Lee, N. Nagaosa and X. G. Wen, Phys. Rev. B \textbf{70}, 214437 (2004).

\bibitem{Lee2008}
S. S. Lee, Phys. Rev. B \textbf{78}, 085129 (2008).

\bibitem{Zhong}
Y. Zhong, K. Liu, Y. Q. Wang and H.-G. Luo, arXiv:cond-mat/1112.6243 (2011).

\bibitem{Vojta}
M. Vojta, J Low Temp Phys \textbf{161}, 203 (2010).

\bibitem{Custers1}
J. Custers et al., Nature (London) \textbf{424}, 524 (2003).

\bibitem{Custers2}
J. Custers, P. Gegenwart, C. Geibel, F. Steglich, P. Coleman and S. Paschen, Phys. Rev. Lett. \textbf{104}, 186402 (2010).

\bibitem{Paul}
I. Paul, C. P\'epin and M. R. Norman, Phys. Rev. Lett. \textbf{98} 026402 (2007).

\bibitem{Pepin}
C. P\'epin, Phys. Rev. Lett. \textbf{98} 206401 (2007).

\bibitem{Pepin2008}
C. P\'epin, Phys. Rev. B \textbf{77}, 245129 (2008).

\bibitem{Paul2008}
I. Paul, C. P\'epin and M. R. Norman, Phys. Rev. B \textbf{78} 035109 (2008).

\bibitem{Senthil2000}
T. Senthil and M. P. A. Fisher, Phys. Rev. B \textbf{62}, 7850 (2000).

\bibitem{Ruegg2012}
A. R\"{u}egg and G. A. Fiete, Phys. Rev. Lett. \textbf{108} 046401 (2012).

\bibitem{Coleman}
P. Coleman, C. P\'epin, Q. Si and R. Ramazashvili, J. Phys.: Condens. Matter \textbf{13}, R723 (2001).

\bibitem{Khodel}
V. A. Khodel, J. W. Clarkb and M. V. Zvereva, JETP Lett. \textbf{90}, 628 (2009).

\bibitem{Continentino}
M. A. Continentino, $Quantum$ $Scaling$ $in$ $Many$-$Body$ $Systems$ (World Scientific Press, Singapore, 2001).

\bibitem{Coleman2005}
P. Coleman, J. B. Marston and A. J. Schofield, Phys. Rev. B \textbf{72}, 245111(2005).

\bibitem{Hackl2008}
A. Hackl and M. Vojta, Phys. Rev. B \textbf{77}, 134439 (2008).

\bibitem{Weber2008}
H. Weber and M. Vojta, Phys. Rev. B \textbf{77}, 125118 (2008).

\bibitem{Vojta2008}
M. Vojta, Phys. Rev. B \textbf{78}, 125109 (2008).

\bibitem{Civelli2010}
I. Paul and M. Civelli, Phys. Rev. B \textbf{81}, 161102(R) (2010).

\bibitem{Kim2010b}
K.-S. Kim and C. P\'epin, J. Phys.: Condens. Matter \textbf{22} 145601 (2010).

\bibitem{Hackl2011}
A. Hackl and R. Thomale, Phys. Rev. B \textbf{83}, 235107 (2011).

\bibitem{Benlagra2011}
A. Benlagra, K.-S. Kim and C. P\'epin, J. Phys.: Condens. Matter \textbf{23} 145601 (2011).

\bibitem{Tran2012}
M.-T. Tran, A. Benlagra, C. P\'epin and K.-S. Kim, Phys. Rev. B \textbf{85}, 165118 (2012).

\bibitem{Kim-K2008}
K.-S. Kim, A. Benlagra, and C. P\'epin, Phys. Rev. Lett. \textbf{101}, 246403 (2008).

\bibitem{Kim-K2009}
K.-S. Kim and C. P\'epin, Phys. Rev. Lett. \textbf{102}, 156404 (2009).

\bibitem{Wen1991}
X. G. Wen, Phys. Rev. B \textbf{44}, 2664 (1991).

\bibitem{Lee2009}
S. S. Lee, Phys. Rev. B \textbf{80}, 165102 (2009).

\bibitem{Metlitski}
M. A. Metlitski and S. Sachdev, Phys. Rev. B \textbf{82}, 075127 (2010).

\bibitem{Mross2010}
D. F. Mross, J. McGreevy, H. Liu and T. Senthil, Phys. Rev. B \textbf{82}, 045121 (2010).

\bibitem{Herzog}
C. P. Herzog, P. Kovtun, S. Sachdev and D. T. Son, Phys. Rev. D \textbf{75}, 085020 (2007)

\bibitem{Lee2009d}
S. S. Lee, Phys. Rev. D \textbf{79}, 086006 (2009).

\bibitem{Sachdev3}
S. Sachdev, Phys. Rev. Lett. \textbf{105}, 151602 (2010).

\bibitem{Sachdev2012}
S. Sachdev, Annu. Rev. Condens. Matter Phys. \textbf{3}, 5.1 (2012).

\bibitem{Hartnoll}
S. A. Hartnoll,  arXiv:hep-th/0903.3246 (2010).

\bibitem{McGreevy}
J. McGreevy, Adv. High Energy Phys. 2010:723105 (2010).

\bibitem{Yu2012}
R. Yu and Q. Si, arXiv:cond-mat/ 1202.6115 (2012).

\bibitem{Stratt}
R. M. Stratt, Phys. Rev. Lett. \textbf{53}, 1305 (1984).


\end{thebibliography}
\end{document}